\documentclass[10pt]{article}
\usepackage{bbm}   
\usepackage{amsthm} 
\usepackage{latexsym} 
\usepackage{epsfig}
\usepackage{color }
\usepackage{amsfonts}
\usepackage{graphicx}

\begin{document}

\newtheorem{assump}{Assumption}
\newtheorem{theorem}{Theorem}[section]
\newtheorem{prop}{Proposition}[section]
\newtheorem{corollary}[prop]{Corollary}
\newtheorem{lemma}[prop]{Lemma}
\newtheorem{definition}{Definition}[section]
\newtheorem{conjecture}{Conjecture}[section]

\newcounter{mnote}
\newcommand\mnote[1]{{\refstepcounter{mnote}%
\marginpar[\itshape\small\raggedleft\themnote.\ #1]%
{\itshape\small\raggedright\themnote.\ #1}}}

\def\endinterproof{\triangle}


\def\bd{\begin{definition}}
\def\ed{\end{definition}}
\def\be{\begin{equation}}
\def\ee{\end{equation}}
\def\bea{\begin{eqnarray}}
\def\eea{\end{eqnarray}}
\def\bitem{\begin{itemize}}
\def\eitem{\end{itemize}}
\def\ben{\begin{enumerate}}
\def\een{\end{enumerate}}
\def\bdescribe{\begin{description}}
\def\edescribe{\end{description}}
\newcommand{\balign}[1]{\begin{alignat}{#1}}
\newcommand{\ealign}{\end{alignat}}


\def\a{\alpha}
\def\b{\beta}
\def\etaperp{{\eta_\perp}}
\def\c{\gamma}
\def\d{\delta}
\def\e{\varepsilon}
\def\l{\lambda}
\def\k{\kappa}
\def\i{\iota}
\def\r{\rho}
\def\s{\sigma}
\def\S{\Sigma}
\def\t{\tau}
\def\tbar{\bar{\t}}
\def\taubar{\bar{\t}}
\def\th{\theta}
\def\om{\omega}
\def\Om{\Omega}
\def\Omtilde{\tilde{\Omega}}
\def\Ups{\Upsilon}
\def\half{\frac{1}{2}}
\def\third{\frac{1}{3}}
\def\quarter{\frac{1}{4}}
\def\P{\Psi}
\def\Pb{\overline{\Psi}}
\def\Ph{\Phi}
\def\M{\textbf{M}}
\def\calA{{\cal{A}}}
\def\ap{{\a'}}
\def\bp{{\b'}}
\def\app{{\a''}}
\def\bpp{{\b''}}
\def\mup{{\mu'}}
\def\nup{{\nu'}}
\newcommand{\ca}{{\underline{\a}}}
\newcommand{\cb}{{\underline{\b}}}
\newcommand{\cc}{{\underline{\c}}}

\def\TD{{\cal{D}}} 
\newcommand{\JT}{J}
\newcommand{\JTtwo}{M}
\newcommand{\JTperp}{N_\perp}
\newcommand{\TCurv}{{\cal{R}}}
\newcommand{\CY}{Y}
\newcommand{\TE}{{\cal{E}}}
\newcommand{\Tg}{{\cal{G}}}
\newcommand{\TC}{{\cal{C}}}
\newcommand{\TB}{{\cal{B}}}
\newcommand{\TA}{{\cal{A}}}
\newcommand{\TAA}{A}
\newcommand{\TZ}{Z}
\newcommand{\cg}{\mathbf{g}}
\newcommand{\ce}{\epsilon}
\newcommand{\co}{\sigma}
\newcommand{\ct}{\tau}
\newcommand{\cu}{\mu}
\newcommand{\cp}{\rho}
\newcommand{\cv}{v}

\newcommand{\Scri}{\mathscr{I}}

\newcommand{\gbar}{\bar{g}}
\newcommand{\ghat}{\hat{g}}
\newcommand{\gtilde}{\tilde{g}}
\newcommand{\hhat}{\hat{h}}
\newcommand{\htilde}{\tilde{h}}
\newcommand{\D}{\nabla}
\newcommand{\Dbar}{\bar{\nabla}}
\newcommand{\Dhat}{\hat{\nabla}}
\newcommand{\Dtilde}{\tilde{\nabla}}
\newcommand{\dv}{\D_v}
\newcommand{\vbar}{\bar{v}}
\newcommand{\bbar}{\bar{b}}
\newcommand{\btilde}{\tilde{b}}
\newcommand{\bhat}{\hat{b}}
\newcommand{\qhat}{\hat{q}}
\newcommand{\qtilde}{\tilde{q}}
\newcommand{\mubar}{\bar{\mu}}
\newcommand{\Ahat}{\hat{A}}
\newcommand{\Atilde}{\tilde{A}} 
\newcommand{\Phat}{\hat{P}}
\newcommand{\Rhat}{\hat{R}}
\newcommand{\Ptilde}{\tilde{P}} 
\newcommand{\Mtilde}{\tilde{M}}
\newcommand{\Rtilde}{\tilde{R}} 
\newcommand{\Etilde}{\tilde{E}} 
\newcommand{\Htilde}{\tilde{H}} 
\newcommand{\etilde}{\tilde{e}}
\newcommand{\utilde}{\tilde{u}}
\newcommand{\thtilde}{\tilde{\theta}} 
\newcommand{\stilde}{\tilde{\sigma}} 
\newcommand{\omtilde}{\tilde{\omega}} 
\def\aveLtilde{\tilde{L}} 
\newcommand{\ttilde}{\tilde{\t}} 
\newcommand{\ntilde}{\tilde{n}} 
\newcommand{\ptilde}{\tilde{p}}
\newcommand{\mutilde}{\tilde{\mu}}
\newcommand{\Ltilde}{\tilde{L}}
\newcommand{\Ombar}{\varphi}
\newcommand{\Upsbar}{\bar{\Ups}}
\newcommand{\oopsbar}{\bar{\Psi}}
\newcommand{\qbar}{\bar{q}}
\newcommand{\den}{\mu} 

\def\Nsc{\textbf{N}}
\def\Asc{\textbf{A}}
\def\Ncf{\textbf{n}}
\def\Acf{\textbf{a}}

\newcommand{\csg}{\mathbf{\epsilon}}
\def\spin{{\cal{S}}}
\def\o{o}
\def\i{\iota}
\def\ot{\tilde{o}}
\def\it{\tilde{\iota}}
\def\O{O}
\def\I{I}
\def\ob{\bar{\o}}
\def\ib{\bar{\i}}
\def\Ob{\bar{\O}}
\def\Ib{\bar{\I}}
\def\ab{\bar{\a}}
\def\dl{D}
\def\dn{\Delta}
\def\dm{\delta}
\def\dmbar{\bar{\delta}}
\def\Mbar{\overline{M}}
\def\mbar{\overline{m}}
\def\Ad{{A'}}
\def\Bd{{B'}}
\def\Cd{{C'}}
\def\Dd{{D'}}
\def\Ed{{E'}}
\def\Fd{{F'}}
\def\Lv{{\cal{L}}}
\def\Nv{{\cal{N}}}
\def\Mv{{\cal{M}}}
\def\Mbarv{{\bar{\cal{M}}}}
\def\zbar{\bar{z}}
\def\Zbar{\bar{Z}}
\def\Dt{\tilde{D}}
\newcommand{\dt}[1]{\frac{\partial {#1}}{\partial t}}
\newcommand{\dr}[1]{\frac{\partial {#1}}{\partial r}}
\newcommand{\dth}[1]{\frac{\partial {#1}}{\partial \th}}
\newcommand{\dph}[1]{\frac{\partial {#1}}{\partial \phi}}
\newcommand{\du}[1]{\frac{\partial {#1}}{\partial u}}
\newcommand{\dxa}[1]{\frac{\partial {#1}}{\partial x^\a}}

\def\rest{\arrowvert} 
\def\implies{\Rightarrow}
\def\iff{\Leftrightarrow}
\def\mod{\arrowvert}
\def\norm{\Vert}
\def\after{\circ}
\newcommand{\phm}{\phantom{-}}  
\def\sect{\in\Gamma}
\def\scri{{\cal{I}}}
\def\Lie{{\cal{L}}}
\def\real{\mathbbm{R}} 

\newcommand{\clos}[1]{\overline{#1}}
\def\Vset{{\cal{V}}}
\def\Uset{{\cal{U}}}
\def\Jset{{\cal{J}}}
\def\Wset{{\cal{W}}}
\def\Xset{{\cal{X}}}

\def\vV{V}
\def\vZ{Z}
\def\vY{Y}
\def\vE{E}

\newcommand{\tensor}[3]{_{#1 \phantom{#2}#3}^{\phantom{#1}#2}}
\newcommand{\abCd}{\tensor{\a\b}{\c}{\d}}
\newcommand{\Abcd}{\tensor{}{\a}{\b\c\d}}

\newcommand{\Sf}[4]{S({#1})_{#2 \phantom{#3}#4}^{\phantom{#2}#3}}
\newcommand{\twovec}[2]{\left(\begin{array}{c}{#1}\\{#2}\end{array}\right)}
\newcommand{\vect}[3]{\left(\begin{array}{c}{#1}\\{#2}\\{#3}\end{array}\right)}
\newcommand{\matrize}[4]{\left(\begin{array}{cc}{#1}&{#2}\\{#3}&{#4}\end{array}\right)}

\def\non{\nonumber}

\def\Rdot{\dot{R}}


\title{A Conformal Extension Theorem based on Null Conformal Geodesics}
\author{Christian L\"ubbe \\ Erwin Schr\"odinger Institute, Vienna, Austria \\ Queen Mary University of London, UK}

\maketitle
\begin{abstract}
In this article we describe the formulation of null geodesics as null conformal geodesics and their description in the tractor formalism. A conformal extension theorem through an isotropic singularity is proven by requiring the boundedness of the tractor curvature and its derivatives to sufficient order along a congruence of null conformal geodesic. This article extends earlier work by Tod and L\"ubbe \cite{article1}.
\end{abstract}

\section{Introduction}

Current observations indicate that on a large scale our universe is almost homogeneous and isotropic and that its evolution is close to the one described by the Friedmann-Lemaitre-Robertson-Walker (FLRW) metrics. These big bang models are characterised by an initial singularity, which under physically reasonable assumptions is also predicted by the singularity theorems of Hawking and Penrose \cite{HEbook}. For the FLRW metrics the initial singularity can be regularised by conformal rescaling as the space-time is conformally flat. What remains unknown is whether for space-times close to FLRW the singularity exhibits this special rescaling property as well.

Following considerations regarding the initial entropy of the gravitational field and its geometrical relation to the Weyl curvature, Penrose conjectured that the Weyl tensor should be finite or even vanish at an initial singularity \cite{PenroseEinsteinCentsurvey}. In order to analyse this Weyl tensor hypothesis (WTH) in a mathematical framework Goode and Wainwright \cite{GoodePWTH, GW} proposed the concept of isotropic singularities, where, similar to the FLRW models, the singularity is a regular hypersurface modulo conformal rescaling. This approach was used by  Anguige and Tod \cite{AT1, AT2} to study the initial value problems for perfect fluid and massless collionless matter with data prescribed on the singularity. Investigating the problem from this end one may ask whether cosmological models with an isotropic singularity evolve close to FLRW.

In \cite{Tod2002}, Tod observed that there exist many different matter models with isotropic singularities and asked whether it was possible to characterise all singularities proposed by the WTH. He outlined in detail how one might be able to use conformally invariant concepts like tractors and conformal geodesics to prove the regularity of the conformal structure at the singularity. It was proposed that finiteness of the conformal curvature in the appropriate sense should lead to the existence of a conformal factor that would regularise the singularity and allow for an extension of the space-time. A recent article \cite{article1} followed this proposal. Given boundedness of the tractor curvature and its derivatives along a congruence of time-like conformal geodesics a local extension theorem for isotropic singularities, referred to as conformal gauge singularities in \cite{article1}, was deduced.

In this paper we will derive an analogous result using null conformal geodesics (Theorem \ref{conformalextthm}). Null conformal geodesics are the same as metric null geodesics as point sets. Thus it is generally easier to find these curves than time-like conformal geodesics and to use them for conformal extension.

\section{Conformal geometry}

We start with a brief summary of some of the important concepts that will be required in this article. For definitions, notation conventions and more detailed analysis of some points the reader is referred to \cite{article1}. We will use $ n \ge 4$  and metric signature $(1,n-1) $ throughout.

Given a non-vanishing section $\s$ of the conformally weighted line bundle $\e[1]$ over a space-time $M$ and a 1-form $b_i$, the equivalence class of pairs $(\s , b_i) \sim (\Om\s, b_i - \frac{\partial_i\Om}{\Om}) $ uniquely determines the connection $\Dhat$ on $\e[w]$ by
$$
  \Dhat_a \t = \s^{w} \partial_a (\s^{-w}\t) + w b_a \t \nonumber \non.
$$
Given a bundle $E$ and connection $\D^E$ we use tensor products and the Leibniz rule to define the connection $\Dhat^E$ on the weighted bundle $E[w]=E\otimes \e[w]$.

The connection $\D$ defined by $(\s,0) $ satisfies $\D_i \s=0$ everywhere.
Conversely, there exists a density $\t$ such that $\Dhat_i \t=0$ if and only if the 1-form $b_i$ is exact.
Any two connections $\Dhat$ and $\D$ are uniquely related by a 1-form $b$, schematically written as $\Dhat = \D + b $, such that for any $\t \sect(\e[w])$ we have
\be
\label{Weylsectiontrans}
  \Dhat_a \t = \D_a \t + w b_a \t
\ee
By identifying the volume density bundle $\e_{[\mu_1,\ldots, \mu_n]}[n]$ of the conformal class $[g] $ with $\e[n]$ one can associate a canonical conformal scale $\s$ to the metric $g_{ij}$ and define the conformal metric $ \cg_{ij}:=\s^2 g_{ij} \sect(\e_{ij}[2]) $ as well as a unique connection $\D $, such that $\D_i \s=0 $ and $\D_k \cg_{ij} = 0$. The conformal metric $ \cg_{ij}$ is defined independently of the chosen metric in $[g]$, while the connection changes as $\Dtilde = \D + \Upsilon $ if one chose $\Omega^2 g_{ij} $ instead.

Any torsion-free connection $\Dhat $ preserving $\cg_{ij}$ gives rise to a general Weyl connection of the conformal class $[g]$ and reduces to a Levi-Civita connection if and only if it preserves a section in $\e[1]$. In light of the tractor formalism used later, connections are referred to as gauge choices.

The connection coefficients of two Weyl connections, $\Dhat$ and $\D$, are related by
\be
\label{Stensor}
\hat{\Gamma}^{k}_{ij}=  \Gamma^{k}_{ij} + S\tensor{ij}{kl}{} b_l, \quad where \quad
S\tensor{ij}{kl}{}=\d\tensor{i}{k}{} \d\tensor{j}{l}{} + \d\tensor{i}{l}{}\d\tensor{j}{k}{} - \cg_{ij}\cg^{kl}
\ee
and $b_l$ is the 1-form in (\ref{Weylsectiontrans}). By applying the Leibniz rule to $\e^i[w] = \e^i \otimes \e[w]$ and $\e_i[w] = \e_i \otimes \e[w]$, equ.(\ref{Stensor}) gives
\bea
  \Dhat_i U^k &=& \D_i U^k + S\tensor{ij}{kl}{} b_l U^j + w b_i U^k , \non \\
  \Dhat_i \omega_j &=&\D_i \omega_j - S\tensor{ij}{kl}{}b_l \omega_k + w b_i \omega_j \non .
\eea
We define weighted spinor sections $\e^A[w], \e^\Ad[w], \e_A[w], \ldots $. In particular we define the weighted spinor metric  $\csg_{AB}:=\s\e_{AB} \sect (\e_{AB}[1])$. Then we get $\cg_{ab}=\csg_{AB}\csg_{\Ad\Bd} $ and $\Dhat_{C\Cd} \csg_{AB} = 0$. \footnote{We could also define $\csg_{AB}=\s^w\e_{AB}$ and $\csg_{\Ad\Bd}=\s^{2-w} \e_{\Ad\Bd}$, without losing consistency for the definition of $\cg_{ab}$. However we then have $\Dhat_{C\Cd}\csg_{AB}=(w-1)b_{C\Cd}\csg_{AB}$.}
We extend the action of general Weyl connections to weighted spinors.
\bea
  \Dhat_{A\Ad}\a^B &=& \D_{A\Ad}\a^B + \d\tensor{A}{B}{}b_{C\Ad}\a^C + w b_{A\Ad}\a^B \non \\
\label{spinorsectiontransform}
  \Dhat_{A\Ad}\a_B &=& \D_{A\Ad}\a_B - b_{B\Ad}\a_A + w b_{A\Ad}\a_B \non
\eea
Given an arbitrary curve $\c$ in $(M,g)$ with tangent vector $v^i$ and a connection $\Dhat=\D + b$ we can define a propagation law for a vector $e^k$ by
\be
\label{Weylprop}
\Dhat_v e^k = \D_v e^k + b_l S\tensor{ij}{kl}{}v^i e^j = 0 \non
\ee
We refer to this transport as $b$-propagation or Weyl propagation. Similarly for spinors we get
\bea
\label{spinorWeylprop}
  \Dhat_v \a^A=0 & \iff & \D_v \a^A = -v^{A\Ad}b_{A\Cd}\a^C ; \\
  \Dhat_v \a_C=0 & \iff & \D_v \a_C = v^{A\Ad}b_{C\Ad}\a_A \non.
\eea
For a general Weyl connection the Schouten tensor is defined as $$ \Phat_{ij} = \frac{1}{n-2} \left( \Rhat_{(ij)} - \frac{n-2}{n}\Rhat_{[ij]} - \frac{1}{2(n-1)}\cg_{ij}\cg^{mn}\Rhat_{mn}\right) .$$ Under $\Dhat = \D + b $ it transforms as
$$ \ P_{ij} - \Phat_{ij} = \D_i b_j - \half b_k b_l S \tensor{ij}{kl}{} .$$
Note that for $\e[w] $ we get $  \Dhat_{[i} \Dhat_{j]} \co = w \Dhat_{[i} b_{j]} \co = - w \co \Phat_{[ij]}$.

 From now on a general Weyl connection will be denoted by $\Dhat$, whereas $\D$ and $\Dtilde$ will denote the Levi-Civita connections of the metrics $g$ and $\gtilde$.

\section{Null conformal geodesics}
A conformal geodesic is a curve $\c(\t)$, with conformal parameter $\t$, velocity vector $v^i$ and 1-form $b_j$, described by the pair of equations
\bea
\label{Cgequ1}
  \D_v v^k + b_l S\tensor{ij}{kl}{}v^i v^j &=& 0 , \\
\label{Cgequ2}
  \D_v b_j - \half b_k b_l S\tensor{ij}{kl}{}v^i &=& P_{ij}v^i ,
\eea
where $P_{ij} $ is the Schouten tensor. The 1-form $b$ induces the general Weyl connection $\Dhat=\D + b$ for which (\ref{Cgequ1}, \ref{Cgequ2}) take the form
\be
\label{Weylcgdef}
  \Dhat_v v^i=0 , \quad \quad\Phat_{ij}v^i=0 .
\ee
%
For a null conformal geodesic (\ref{Cgequ1}) takes the form $\D_v v^i + 2 \langle b,v \rangle v^i = 0 $, where $\langle b,v \rangle = b_k v^k $. Thus any null conformal geodesic is a metric null geodesic, though not necessarily affinely parametrised. The converse is shown by the following proposition.

\begin{prop}
Suppose we are given an affinely parametrised metric null geodesic $\c(s) $ with velocity vector $u^i$ and a solution for the following system of differential equations along $\c(s) $
\bea
\label{cge2null}
  \D_u b_j  &=& P_{ij}u^i + b_k u^k b_j - \half g_{ij}u^i g^{kl}b_k b_l  \\
  - q^{-1}\frac{dq}{ds}&=& \langle b,u \rangle .
\eea
It follows that
\be
\label{cgeqnull}
  \frac{d^2 q}{ds^2} = -P(u,u) q
\ee
Furthermore in the orientation preserving reparameterisation
\be
\label{taudefinition}
  \t(s) :=  \int^s q^{-2} ds'.
\ee
$\c(\t)$ is a conformal geodesic with velocity $v^i=q^2 u^i $ and 1-form $b_j$.
\end{prop}
We note that an orientation reversing parametrisation can easily be obtained by setting $\t(s) := - \int^s q^{-2} ds'$ and $v^i=-q^2 u^i $.
\begin{proof}
Observe that (\ref{Cgequ2}) is linear in $v^i$. We begin by solving (\ref{Cgequ2}), with $u^i$ replacing $v^i$, for the 1-form $b_i(s)$ along the curve. This gives (\ref{cge2null}), where the tensor $S\tensor{ij}{kl}{} $ has been expanded.
The contraction of (\ref{cge2null}) with $u^j$ together with $\D_u u^i=0 $ and $\langle b,u \rangle = - q^{-1}\frac{dq}{ds}$ then gives (\ref{cgeqnull}). The velocity $v^i=q^2 u^i $ and the conformal parameter (\ref{taudefinition}) satisfy (\ref{Cgequ1}). The reader should observe that the freedom in the initial data for $q$ reflects the freedom to change the conformal parameter $\t$ by a fractional linear transformation \cite{Friedrich, article1}.
\end{proof}
Any solution of (\ref{cge2null}, \ref{cgeqnull}) leads to a solution of (\ref{Cgequ1}, \ref{Cgequ2}) and defines an associated Weyl connection $\Dhat = \D + b $ along $\c$ for which (\ref{Weylcgdef}) holds.
We use a conformal factor $\Om$ on $M$ with $\Om\rest_{\c} := q^{-1} $ to define the metric $\gtilde_{ij}=\Om^2 g_{ij}$ and its Levi-Civita connection $\Dtilde=\D + \Ups $, where $\Ups=\frac{\partial_i \Omega}{\Omega} $. For $\Dtilde$, $\c$ is a metric null geodesic affinely parametrised by $\t$, as well as a conformal geodesic with 1-form $\btilde=b-\Ups $ such that $\langle \btilde ,v \rangle = 0$. This shows that we can refer to the curves as null conformal geodesics or null metric geodesics alike. In the following we will consider smooth congruences of conformal geodesics with a smooth conformal factor $\Om$ chosen such that $\Om := q^{-1} $ for every conformal geodesic.

For notational convenience we will denote the derivatives along $\c$ by $D=\D_u=\frac{d}{ds} $ and $\Dt=\D_v=\frac{d}{d\t} $.

\section{Spinors}
To analyse the properties of null conformal geodesics we start with a metric geodesic $\c(s) $. We choose a spinor dyad $\{\o^A, \i^A\} $ normalised with $\e_{AB} $ and $\D $-propagated along  $\c(s) $ such that $u^a=\o^A\o^\Ad $. We solve (\ref{cgeqnull}) for $q$ and define a second dyad $\{\ot^A, \it^A\} = \{q\o^A, \i^A\} $ normalised with $\tilde{\e}_{AB} = q^{-1}\e_{AB}$. The indices of each dyad are lowered with the corresponding spinor metric, i.e. $\{\ot_A, \it_A\} = \{\o_A, q^{-1}\i_A\} $.

In the following we solve for the 1-form $b$ on $\c$ in terms of the two dyads and analyse the family of solutions in more detail. We also find an expression for all Weyl propagated spinors along $\c$ in terms of the dyads, as these will be used later to discuss the boundedness of curvature.

Both dyads have an associated Newman-Penrose tetrad (NP tetrad) which are denoted $\{u^a, w^a, z^a, \zbar^a\}$ and $\{l^a, n^a, m^a ,\mbar^a \}=\{q^2 u^a, w^a, qz^a, q\zbar^a\}$. Note that the first tetrad is $\D$-propagated, whereas only $l^a=v^a$ is $\Dtilde$-propagated for sure. We expand $b_i$ in the dual NP tetrad again using a different metric for each tetrad.
\be
 b_c=Wu_c+Uw_c-\Zbar z_c-Z\zbar_c =Nl_c +Ln_c - \Mbar m_c - M \mbar_c   \nonumber
\ee
where $N=W, \, L=q^2 U, \, M=Zq $. Substituting into (\ref{cge2null}) gives
\be
\begin{array}{lclclcl}
  DU-U^2&=&P(u,u)&\quad\quad &\Dt L + L^2 &=&P(l,l) ;\\
  D(Zq)&=&P(u,z)q & \quad\quad &  \Dt M&=&P(l,m) ; \\
  DW - Z \Zbar &=&P(u,w) & \quad\quad & \Dt N - M \Mbar &=&P(l,n).
\end{array} \nonumber
\ee
Using $U=-\frac{Dq}{q}$ we get $D^2q =-P(u,u)q $. Similarly $L=\frac{\Dt \phi}{\phi} $ gives $\Dt^2 \phi = P(l,l)\phi $. Furthermore $\Dt=q^2 D $ and $q^2 U = L $ imply that $q\phi$ is constant along $\c$. So $\phi=A \Om $ for some constant $A$. The other components of $b$ are found by integration along $\c$.

Since $l^a= \ot^A \ot^\Ad $ is Weyl propagated it follows that $\ot^A $ solves (\ref{spinorWeylprop}). The spinor metric $\tilde{\e}_{AB} $ is also Weyl propagated along $\c$. Hence the other Weyl-propagated spinor in the normalised (with respect to $\tilde{\e}_{AB} $)  dyad has to be of the form $\a^A = f\ot^A + \i^A$ for some function $f$ along $\c$. Substituting $$  \a^A= \tilde{X}\ot^A +\tilde{Y}\it^A = X \o^A + Y \i^A$$ into  (\ref{spinorWeylprop}) gives $q\tilde{X}=X, \tilde{Y}=Y $ and
\be
  \Dt\tilde{X} = -\Mbar \tilde{Y}  , \quad \Dt \tilde{Y} = 0 \nonumber
\ee
This has solution $$ \tilde{X}= - \tilde{Y}_* [\Mbar_*\t + \int^\t_0 \int^{\t'}_0 P(l,\mbar)d\t'd\t'' ] + \tilde{X}_*, \quad \tilde{Y}=\tilde{Y}_* ,$$ where a subscript $*$ denotes initial data. The initial dyad $\{q_*\o^A, \i^A \}$ gives rise to the solution $\{\O^A, \I^A\} $ which forms a basis for all Weyl propagated spinors along $\c$.
The transformation between $\{\o^A, \i^A\} $ and $\{\O^A, \I^A\} $ can be expressed as
\bea
\label{spinormatrix}
  \twovec{\O^B}{\I^B}&=&\matrize{1}{0}{f}{1} \matrize{q^\half}{0}{0}{q^{-\half}} q^\half \twovec{\o^B}{\i^B} ,
\eea
where $f(\t)= -\Mbar_*\t - \int^\t_0 \int^{\t'}_0 P(l,\mbar)d\t'd\t'' $.
Thus we have a combination of a null rotation, a boost and a rescaling. The NP tetrad $\{\hat{l}^a, \hat{n}^a, \hat{m}^a ,\hat{\mbar}^a \}$ formed from $\{\O^A, \I^A\} $ is a null rotation along $\c$ of the tetrad $\{l^a, n^a, m^a ,\mbar^a \}$ depending on $P_{ij} $ and $\t $. 


We define the section $v:=\csg_{AB}\O^A \I^B$ as our preferred conformal scale and observe that $v^{-2}\cg_{ij} = \gtilde_{ij} = q^{-2} g_{ij} $ with $\gtilde(n,l)=1 $ and $\Dhat_v v =0 $. This generalises the definition that we gave for the time-like case \cite{article1}. We will refer to $v$ and $\Dtilde = \D^{\gtilde} $ as the \textit{normal scale} and the \textit{normal gauge}.

Let us consider a vacuum space-time ($P_{ij}=0$) for a moment. The solutions are given by, ($q_1 $ is a constant),
\bea
\label{qvacuum}
  q=q_1 s+ q_*, \quad \t-\t_*= \frac{s}{q_* q} \\
\label{bvacuum}
  U=-\frac{q_1}{q}, \quad W=Z_* \Zbar_* \frac{q_* s}{q} + W_*, \quad Z=Z_*\frac{q_* }{q} \\
\label{vacuumspinormatrix}
  \twovec{\O^B}{\I^B} = \matrize{q_1 s + q_*}{0}{-\Zbar_* s}{1}\twovec{\o^B}{\i^B}.
\eea
Remark: If a cosmological constant $\lambda $ is introduced or equivalently an Einstein metric ($P_{ij}=\quarter P g_{ij}$) is used, then the solution is the same up to an additional term of $\quarter P s $ for $W$.

We rewrite (\ref{qvacuum}) as $\t = \frac{s}{q_* q_1(s-s_\infty)} $, where $ s_\infty = -\frac{q_*}{q_1} = \frac{1}{U_*}$ (with $ s_\infty =\pm \infty$ for $q_1=0$). This shows that the conformal parameter is fixed up to a M\"obius transformation induced by the initial data for $W_* $. At $s=s_\infty$, $\t $ passes through $\infty$ and enters a new `cycle'. We refer to these points between the cycles as poles, as $\t$ diverges with respect to $s$. Furthermore $q$ and $ v^i$ vanish at $s_\infty $, while $b_i$ has a pole. If $s_\infty < s_0 $, then the components of $b_i$ for vacuum, namely $U, W, Z, \Zbar $ in (\ref{bvacuum}), are bounded with respect to $\{\o^A, \i^A\} $ on $[s_0,\infty) $.

Let $(M,g)$ be a general space-time with a null geodesic $\c$. Suppose we have two solutions of (\ref{cge2null}, \ref{cgeqnull}) denoted by hatted and barred quantities. Their associated Weyl connections satisfy $\Dhat = \bar{\D} +b $ with $b=\bar{b}-\bhat$. In the $\Dhat$-gauge we have $\Phat_{ij}v^i = 0 $, so that $b$ satisfies the vacuum equations (\ref{qvacuum}, \ref{bvacuum}) with the appropriate substitutions $s \to \t $, etc. Thus we have $b=b_{vac} $ and hence $\bar{b} = \bhat + b_{vac}(\hat{\t})$ in the appropriate sense. Furthermore $ \bar{q} = \qhat (q_1 \hat{\t} + q_*), \bar{\t}-\bar{\t}_*= \frac{\hat{\t}}{q_* q_1(\hat{\t}-\hat{\t}_\infty)} $. The associated spinor dyads are related by (\ref{vacuumspinormatrix}).


Given a null geodesic $\c$ two solutions to the conformal geodesic equations are related by a vacuum solution as detailed above. We can see that these solutions are defined on different segments of $\c$ and related by a M\"obius transformation on the overlap. Furthermore we can extend each solution onto the full segment covered by the other one, by demanding that there the solutions be related by exactly the same vacuum solution $b_{vac}$. Repeating this process we are able to extend the solution through the point where $\t$ and $b$ have a pole and hence treating $\t$ as a projective parameter now. Alternatively the M\"obius transformations allow us to shift the pole in $\t$ and $b$ by changing the initial data, while the curve remains fixed as a point set. An important quantity is the overall number of cycles between the poles of $\t$  along $\c$. It can at most change by one under a M\"obius transformation. Hence to say that a segment of $\c$, in our case the final one, has a finite number of cycles to the future is a conformally invariant statement, while an infinite number of cycles to the future (or the past) gives a conformally invariant notion of infinitely far away. As the example of the Einstein static cylinder in \cite{article1} shows one cannot expect a conformal extension if in the latter case.

As we have seen  we cannot say that a null conformal geodesic is complete if it is defined for all values of $\t$, since this statement depends on the solution. Instead we give the following conformally invariant definition.
\bd
A conformal geodesic is said to be incomplete if there exists a conformal parameter $\t$ for which it is incomplete.
\ed
We can thus see that a conformal geodesic is complete if and only if it has an infinite number of cycles in $\t$ to the future and the past of any point on it. We can now state the following useful lemma
\begin{lemma}
\label{normequiv} Suppose we are given a metric null geodesic $\c $ in $(M,g) $ with two different solutions, $(b,q,\t)$ and $(\bar{b}, \bar{q}, \bar{\t}) $ and associated Weyl propagated NP tetrads (\ref{Weylprop}). Let $M: \t \to \bar{\t}$ denote the M\"obius transformation between $\t $ and $\bar{\t} $. Suppose further that we have an interval $[\t_0, \t_1] $ on which $M $ is continuous and hence $\bar{\t} $ does not diverge. Then the two Euclidean norms calculated from the frame components of any tensor $T$ are equivalent where defined on $[\t_0, \t_1] $. Furthermore if $\c $ is incomplete in $\t$ ending at $\t_F \in [\t_0, \t_1]  $, then $\c $ is also incomplete in $\bar{\t} $ ending at $\bar{\t}_F = M(\t_F) $.
\end{lemma}
Remark: Note that the M\"obius transformation $M$ is still well defined outside the domain of $\c$, in particular if one views $M$ as a map on $S^1 \simeq \real \cup \{\infty\}$

For the extension theorem we will use an incomplete null conformal geodesic such that $\t $ is finite on the final segment, from here on denoted $[t_0, t_F) $.

At every pole $q$ vanishes. However if $q$ vanishes at a singularity then $\t$ can be finite on the final segment of $\c$. For example for $P(u,u)=\frac{2}{9s^2} $ along $\c$ we have $q = s^\third =\t-\t_*$ as a possible solution, where the vanishing of $q$ clearly doesn't coincide with a pole in $\t$ at the ideal endpoint.


\section{Tractors}
For details and notation for the tractor formalism the reader is referred to \cite{BEG, article1}.
From here on let $\c$ be a null conformal geodesic with velocity $v^i=l^i$, Weyl propagated NP tetrad $\{l^i, n^i, m^i, \mbar^i\} $ and induced normal scale $\cv $. Define the tractor $Z^I= v^{-1}X^I $, where $X_I=(1,0,0)$ is the canonical null tractor of weight $w=1$. Define the velocity and acceleration tractors $  V^I = \TD_v Z^I, \,  A^I = \TD_v V^I$, where $\TD$ denotes the tractor connection.
Using the Weyl connection associated to the null geodesic as a tractor gauge, indicated as a subscript, we can see that
\be
  V^I=L^I=\vect{0}{v^{-1}l^i}{0}_{\Dhat} \, , \quad A^I=0. \non
\ee
In the Weyl gauge $\Dhat$ the tractors $E_I =(0, \,v^{-1}\cg_{ij}e^j ,\,0))$ will represent Weyl propagated vectors $e^i $ iff $\TD_v E_I = v^{-2}\cg(e,v) (0, 0 ,-v)  $. Hence $\TD_v M^I=0$, $\TD_v \Mbar^I=0,\, \TD_v L^I=0$, while
\be
\label{Bnull}
 \TD_v N^I = B^I = \vect{-\cv}{0}{0}_{\Dhat} \, \mathrm{and} \, \TD_v B^I = 0 \non
\ee
Thus $B^I$ plays the same role as the acceleration tractor $A^I$ did in the time-like case \cite{article1}. It is a null tractor and for a different gauge choice it will depend only on the associated 1-form $b$ in that gauge and the conformal density $v$.

For our calculations we will use two orthogonal space-like unit vectors $e^i_{2/3}$ that are given by a constant linear combination of $m^i, \mbar^i $. These are also Weyl propagated and the corresponding tractors satisfy $\TD_v E_\app=0 $, where the index $\app$ takes values 2 or 3 and the summation over these two values applies for repeated indices. We denote the set $\{l^i, n^i, e^i_2, e^i_3 \}$ by $\{e^i_\a\} $ and the corresponding tractors by $\{E^I_\a\} $. These then form the tractor basis $\{\TE_\TB^I\}=\{Z^I, L^I, N^I, E_1^I, E_2^I, B^I \}$.

\begin{prop}
\label{Normequiv} Suppose we are given conditions as in proposition \ref{normequiv}. Then on $[\t_0, \t_1] $ the Euclidean norms calculated from the frame components of any tractor $T$ are equivalent.
\end{prop}
\begin{proof}
Note first that the two normal scales $\cv$, $\bar{\cv}$ are related by $q=q_1 \t + q_* $, which can't vanish on $[\t_0, \t_1] $ otherwise $\bar{\t} $ would have a pole there. 
We choose the Weyl gauge of the first solution to express the tractor basis associated to the second solution. So below $b$ is the vacuum 1-form solution relating the initial solutions. We have
\be
\bar{B}^I = \vect{-\bar{\cv}}{\bar{\cv} \cg^{ij}b_j}{\half \bar{\cv}\cg^{ij}b_i b_j}_{\Dhat} \, , \,
\bar{E}^I_\a=\vect{o}{\bar{\cv}^{-1} e^i_\a}{\bar{\cv}^{-1} b_i e^i_\a}_{\Dhat} \, , \, q\Zbar^I= Z^I \non
\ee
It follows that the frame components of the second tractor frame are bounded under the conditions of the corollary. These conditions also imply that $\t$ doesn't have a pole in $\bar{\t}$ on the above segment of $\c$. Since the two norms are related continuously the result follows.
\end{proof}
The tractor curvature $\TCurv\tensor{ij}{K}{L}$ is defined by $ \left( \TD_i \TD_j - \TD_j \TD_i \right)U^K = \TCurv\tensor{ij}{K}{L} U^L$. In the remainder of this article we will derive conformal extension theorems from bounds on the tractor curvature and its derivatives. Under the conditions in propositions \ref{normequiv} and \ref{Normequiv} we have a well-defined notion of the finiteness of tractor curvature.

\section{Construction of a null congruence and conformal Gaussian coordinates}
\label{nullcoord}
Assume we are given an incomplete conformal geodesic $\c$ in $(M,g)$. Choose a point $p$ on a final segment of $\c$ where the conformal parameter $\t$ is finite. Choose a spinor dyad $\{\o^A, \i^A\} $ at $p$ and parallelly propagate it along $\c$ generating an NP tetrad $\{u^a, w^a, z^a, \zbar^a\} $ so that $u^a=\o^A \o^\Ad$ is the velocity of the metric geodesic.

We first construct a metric geodesic null congruence $C$ \cite{PenroseSIAM7,Racz}. Let $Q$ be a sufficiently small neighbourhood of the origin in $span\{z, \zbar\}\subset T_p M$ and define the 2-surface $\Lambda:=exp[Q] $ at $p$. This setup is useful for our later analysis, however any suitable 2-surface $\Lambda$ can be chosen. The null vectors $u^a, w^a $ can be extended smoothly across $\Lambda$.
Let $C_1$ be the congruence of null geodesics along $u^a$, and hence containing $\c$, and $C_2$ the congruence of null geodesics along $w^a$. The congruences $C_1$ and $C_2$ generate null hypersurfaces $\Pi $ and $\Sigma$ with normal vectors $u^a $ and $w^a$ respectively.
Moving $\Lambda$ along $C_2 $ we can generate a 1-parameter family of null hypersurfaces $\{\Pi\}_u$, with $u=0$ giving $\Pi$. This generates a 3-parameter family $C$ of metric null geodesics with a smooth tangent vector field $l$ in a neighbourhood of $\Sigma $ and each curve $\d\in C $ transversal to $\Sigma$. Extend the spinor dyad  $\{\o^A, \i^A\} $ smoothly across $\Sigma $ and parallelly propagate it along $C$. Let $x_{\ca '} $ be coordinates on the closure $S$ of a simply connected open neighbourhood of $p$ in $\Sigma$ with origin at $p$.

For the conformal geodesics we specify smooth initial data for $q $ and $b_i$ across $\Sigma$ and solve (\ref{cge2null}, \ref{cgeqnull}). We then solve (\ref{taudefinition}) with $\t$ synchronized with respect to $\Sigma$ and develop the Weyl propagated frame and the associated tractor basis to measure the tractor curvature. As we have seen boundedness of $\TCurv\tensor{ij}{K}{L}$ along $C$ depends on the initial value $L_*=-\frac{(Dq)_*}{q_*}$, as this can change the location of the poles of $\t$. The rest of the initial data is freely specifiable. In particular initial data $q=1, b_i=0$ on $\Sigma$ we have $\gtilde_{ij} = g_{ij}$ and $\hat{\Gamma}^{k}_{ij} = \tilde{\Gamma}^{k}_{ij} = \Gamma^{k}_{ij}$ on $\Sigma$.
\begin{assump}
\label{Uassump}
In the following we are given $U\subset M$ with these properties:
\bitem
\item $U$ contains a final segment of $\c$
\item $U$ is closed in $M$ and strongly causal
\item $U$ has an open neighbourhood $U'$ in $M$ such that if $\d \in C$ crosses $S$ then $U' \cap \d $ is connected with $\t \ge 0$ and bounded above on $U' \cap \d $.
\item any two elements of $C$ are disjoint in $U'$.
\eitem
\end{assump}
The coordinates $x_{\ca '} $ on $S$ are dragged along $C$ and together with the conformal parameter $\t $ generate conformal Gaussian coordinates $\{\t, x_1, x_2, x_3\} $ \cite{FS} on $U$. Hence elements of $C$ are given by lines of constant $x_{\ca '} $ and $\t =0 $ on $S$.

Let $\phi$ be the coordinate chart of a neighbourhood $U'$ of $U$ associated to the conformal Gaussian coordinates and denote its inverse by $\psi$. We denote the images of $U'$ and $U$ by $V'$ and $V$ respectively. Note that $V=\phi(U)$ is not closed in $\real^n$ since $\c(\t)$ is incomplete. Our aim will be to extend functions (these will be the coordinate components of the unphysical metric) onto an open neighbourhood $\tilde{V}$ of $V$ in $\real^n$ and thus find an extension on a neighbourhood $\tilde{U}$ of $U$ provided the conditions of the extension theorems given below are satisfied. The reader should note that the functions on $\tilde{U}$ and $U'$ need not agree in value, in which case we only get a local extension.

In order for the coordinates to work on the extension one has to prevent the crossing of the conformal geodesics or the development of conjugate points in $\tilde{U}$. The former can be guaranteed using the fourth condition on $U$ above. For the latter it will be shown that there exists $\epsilon>0$ such that the norm of the conformal Jacobi fields associated to $C$ is bounded below by $\epsilon$ in $U$. More details are given below in section \ref{CJE}. In section \ref{sectcjp} we will deduce the existence of $\epsilon$ from bounded tractor curvature.

\section{The conformal Jacobi equation}
\label{CJE}

Given a congruence of null conformal geodesics with smoothly varying velocity and 1-form fields $v^i$ and $b_j$ and deviation vector $\eta^i$, one can derive the conformal Jacobi equations \cite{Friedrich}
\bea
\label{confJac1}
  \D_v^2 \eta ^k &=& R\tensor{ij}{k}{l}v^i \eta ^j v^l - \D_\eta \left(b_j S\tensor{il}{jk}{}v^i v^l \right) , \\
\label{confJac2}
  \D_v \D_\eta b_l &=& -b_k R\tensor{ij}{k}{l} v^i \eta ^j + \D_\eta (P_{kl}v^k) + \half \D_\eta \left(b_j b_k S\tensor{il}{jk}{}v^i \right) .
\eea
In the Weyl connection $\Dhat=\D+b$ the first equation takes the form
\be
\label{confJachat}
  \Dhat_v^2 \eta^k = \Rhat\tensor{ij}{k}{l}v^i \eta ^j v^l . \non
\ee
The second equation becomes an identity. If we define the Jacobi tractor $\JT^I:=\TD_\eta Z^I$ then (\ref{confJac1}, \ref{confJac2}) can be rewritten in terms of tractors as $$\TD_v ^2 \JT^K = \TCurv(v,\eta)\tensor{}{K}{L}V^L + \TD_\eta A^K .$$ Since the acceleration tractor $A^I$ vanishes in the null case it follows that
\be
\label{tractorconfJac1}
  \TD_v ^2 \JT^K = \TCurv(v,\eta)\tensor{}{K}{L}V^L.
\ee
This leads to a number of simplifications for the analysis of the null conformal geodesics compared with that of timelike ones \cite{article1}.

In order to generalise (\ref{tractorconfJac1}) we introduce the following notation for the $k^{th}$ coordinate derivative of the Jacobi tractor. $Y^I_{(k)}=\TD_\xi^k \JT^I :=\TD_{\xi_k} \ldots \TD_{\xi_1} \JT^I$. We set $y^\TB_{(k)}:=\partial^k_\xi (\JT^I \theta_I^\TB)$, where $\{ \theta_I^\TB \}$ is the dual tractor basis. Note that $\JT^I \theta_I^\b = \gtilde(\eta, e_\a)\eta^{\a\b} $, so that we can use the same notation for the Jacobi vectors. Then the generalised Jacobi equation for tractors is given by:
\be
\label{generalconfJac}
  \TD_v ^2 Y_{(k)}^K = \TCurv(v,e_\b)\tensor{}{K}{L}V^L y^\b_{(k)} + Q_{(k)}^K ,
\ee
where $Q_{(k)}^K$ is defined recursively by
\be
  Q_{(0)}^K=0 ,
\ee
\bea
\label{Qkformula}
 Q_{(k)}^K &=& \TD_v (\TCurv (v, \xi )\tensor{}{K}{L} Y_{(k-1)}^L )  + \TCurv (v, \xi )\tensor{}{K}{L}\TD_v Y_{(k-1)}^L  \non \\
& & + \TD_\xi (\TCurv(v,e_\b)\tensor{}{K}{L}V^L)  y^\b_{(k-1)} + \TD_\xi Q_{(k-1)}^K .
\eea
This is the null version of proposition 4.3. in \cite{article1}. The timelike analogue of (\ref{generalconfJac}) is a third order equation. However due to the vanishing of the acceleration tractor (\ref{generalconfJac}) is second order for a null conformal geodesic and this simplifies the formula for $ Q_{(k)}^K $. This version of (\ref{generalconfJac}-\ref{Qkformula}) is similar to the tensor analogue in \cite{Racz}.

For our calculations we introduce:
\be
 \label{defTTbar}
  T^I := y^\TB_{(k)} \TE^I_\TB , \quad  \bar{T}^I :=Y^I_{(k)}-T^I, \quad \bar{Q}^I :=\TD^2_v \bar{T}^I \non
\ee
The expansion of $\bar{T}^I $ contains $y^\TB_{(l)}, \, l<k,$ and $\TD^m \TE^I_\TB, \, m \le k $, so that $\bar{Q} $ contains their time derivatives up to second order. 
The extension theorem requires us to prove the boundedness of $y^\TB_{(k)}$. We use (\ref{generalconfJac}) to formulate a comparison theorem (proposition \ref{Upperboundtheorem}). To simplify the expressions that we shall obtain, we drop the subscript $(k)$.
We also define
\bea
  A_{\a\b} = \TCurv(v, e_\b)( E_\a,V)=C_{ijkl}v^i e^j_\b e^k_\a v^l , \nonumber \\
  C_{\b} =  \TCurv(v, e_\b)(V,B)=\widehat{\CY}_{ijk}v^i e^j_\b v^k , \nonumber
\eea
where $\TCurv(v,w)(X,Y)=\TCurv_{ijKL}v^i w^j X^K Y^L $. Both vanish when an index is zero. Substitute $\TD^2_v Y^I=\TD^2_v T^I+\bar{Q}^I $ into (\ref{generalconfJac}), then it splits into the following hierarchy of differential equations for $y^\TB_{(k)} $. 
\bea
\label{JacobiL}
  D^2 y_0 &=& (Q - \bar{Q})_0 , \\
\label{JacobiZ}
  D^2 y_\TZ &=&  - 2 D y_0 + (Q - \bar{Q})_Z  , \\
\label{Jacobiapp}
  D^2 y_\app &=& A_{\app\b}y^\b + (Q - \bar{Q})_\app ,\\
\label{JacobiB}
  D^2 y_B &=& - C_\a y^\a + (Q - \bar{Q})_B ,\\
\label{JacobiN}
  D^2 y_1 &=& A_{1\b}y^\b -  2 D y_B + (Q - \bar{Q})_1.
\eea
We can use this system of equations to prove the following theorem:
\begin{prop}
\label{Upperboundtheorem}

Suppose the following hold for $U$ as defined above:
\newline i)~ there exists a bound $\TCurv_0$ for the norm of $\TCurv$;
\newline ii) there exists a bound $\half q_0 $ for the norms of $Q^{(k)}$, $\bar{Q}^{(k)}$.
\newline Then $y^\TB_{(k)} = \TD^k_\xi (\JT^\TB)$ and its two time derivatives are bounded over U.
\end{prop}
\begin{proof} We prove the required bounds for an arbitrary curve $\d \in C $ and hence for $U$. We start by integrating (\ref{JacobiL}) twice and deduce that $y_0 $ is bounded on any finite interval. For later use denote this bound by $\l$. Substituting this into (\ref{JacobiZ}) gives us the boundedness of $y_Z$ and its two time derivatives.

For (\ref{Jacobiapp}) we recall that $A_{\app\b} $ vanishes for $\b=0$ and that $y^1=y_0$. Hence we can rewrite (\ref{Jacobiapp}) as
\be
\label{Jacobiapp2}
  D^2 y_\app = A_{\app\bpp}y^\bpp +  A_{\app1} y_0 + (Q - \bar{Q})_\app
\ee
We define $y^i=y^\app e^i_\app $ and $z=\norm y^i \norm$. From (\ref{Jacobiapp2}) we derive
\bea
  & & \norm y_\app (\t) \norm - \norm y_\app(0)\norm - \norm \t Dy_\app(0) \norm\non \\
  &\le&  \norm y_\app (\t) - y_\app(0) - \t Dy_\app(0) \norm \non \\
  &=& \norm \int^\t_0 \int^\s_0 (A_{\app\bpp}y^\bpp +  A_{\app1} y_0 + (Q - \bar{Q})_\app )(\s') d\s' d\s \norm \non\\
\label{yappinequality}
&\le &  \int^\t_0 \int^\s_0 (\TCurv_0 z(\s') + \TCurv_0 \l + q_0) d\s' d\s
\eea
Now let $y_\e(\t)$, $(\e>0)$, be a solution of
\be
\label{ymaster}
  y_\e(\t)-y_\e(0) - \t Dy_\e(0) = \int^\t_0 \int^\s_0 (\TCurv_0 y_\e(\s') + \TCurv_0 \l + q_0) d\s' d\s
\ee
with initial data $ y_\e(0)=z(0) + \e, Dy_\e(0) = \norm Dy_\app \norm_0$. Define
$$\t_{max}:=sup_{\t \in [0,T]} \{\t : z(\s) \le y_\e(\s) \,\,  \forall \s \in [0,\t]\} .$$
Subtracting (\ref{ymaster}) from (\ref{yappinequality}) gives
\be
  z(\t) - y_\e(\t) + \e \le \int^\t_0 \int^\s_0 \TCurv_0 (z-y_\e)(\s') d\s' d\s . \non
\ee
By continuity $z \le y_\e $ for some $\t \in [\t_{max}, \t_1]$, which leads to a contradiction unless $\t_{max}=\t_1$. Thus for any $\e>0$, $z$ is bounded by $y_\e$ on $[0,\t_1]$. Now taking the limit $\e \to 0$ we see that there exists a solution $y_0$ of (\ref{ymaster}) with initial data $ y_0(0)=z(0), Dy_0(0) = \norm Dy_\app \norm_0$. and $z(\t) \le y(\t) \phm \forall \t \in [0, T]$.

For equation (\ref{JacobiB}) the right hand side is now bounded, so that we can deduce the boundedness of $y_B$ by integrating twice. Analogously it follows that  $y_1$ is bounded, which concludes the proof.
\end{proof}
Above we have assumed the existence of the bound $\half q_0$. We now show that if we are given suitable boundedness of the tractor curvature, we can deduce that $Q_{(k)} $ and $\bar{Q}_{(k)} $ are bounded and hence that $q_0$ exists. The proof requires a long induction process and is essentially the same as in \cite{article1}. Since the setup in \cite{article1} uses abstract frame indices most proofs of the required lemmas follow the same pattern. Therefore we will only give a brief summary here, highlighting the main steps.

For the $k^{th}$ induction step it is assumed that for $l \le k $, $m<k$ and $q\le2$ the following bounds are given:
\be
  \TD^l_e \TCurv\tensor{\mu\nu}{K}{L}, \quad \TD_v^q\TD^m \TE^I_\TB, \quad \TD_v^q (y^\TB_{(m)}). \non
\ee
Looking at the decomposition of $Q_{(k)} $ we can see that we require bounded curvature derivatives up to order $k$ and bounds on $Y_{(l)}^L $ and $y^\b_{(l)} $ for $l<k$.
For $\bar{Q}_{(k)}$ it was shown earlier that we require bounded second time derivative of $y^\TB_{(l)}, \, l<k,$ and $\TD^m \TE^I_\TB, \, m \le k $.
Furthermore it is outlined in \cite{article1} that $\TD^l_\xi \TCurv\tensor{\mu\nu}{K}{L} $ can be expressed in terms of $ \TD^l_e \TCurv\tensor{\mu\nu}{K}{L}$ and $y^\TB_{(l)}, \, l<k,$. The proofs make use of the integration lemmas 4.1. and 4.2. as well as equation (62) for permuting derivatives (labels as in \cite{article1}).

Then the steps are as follows (for details see \cite{article1})
Firstly observe that the conditions are sufficient to prove that $ \TD^l_\xi \TCurv\tensor{\mu\nu}{K}{L}$ have bounded components.
By swapping the time derivatives past all the other ones in $\TD_v \TD_\xi^k \TE^I_\TB $ extra tractor curvature terms up to order $k-1$ are introduced, which are all bounded. Thus $ \TD_\xi^k \TE^I_\TB $ is bounded.
It follows that $Q_{(k)} $ and $\bar{Q}_{(k)}$ are bounded and Proposition \ref{Upperboundtheorem} applies.

We can hence reformulate Proposition \ref{Upperboundtheorem} and deduce the boundedness of the components of the unphysical metric.
\begin{prop}
\label{bounded metric components}
Suppose that for $l \le k $ the tractor curvature terms $\TD^l_e \TCurv\tensor{\mu\nu}{K}{L}$ have bounded norms in $U$.
Then $y^\TB_{(k)} = \TD^k_\xi (\JT^\TB) $ and $\TD_\xi^k \JT^I_\ca $ as well as their second time derivatives have bounded norm in $U$.
Furthermore the coordinate components of the unphysical metric $ \gtilde_{ij} $ and its derivatives to order $k$ are bounded.
\end{prop}
\begin{proof}
The boundedness of $\norm \TD_\xi^k \JT_\ca \norm$ follows from the decomposition into $y^\TB_{(l)}$ and $ \TD_\xi^l \TE^I_\TB $ for $l \le k $. For the unphysical metric we observe
\be
\label{unphysical metric}
 \gtilde_{\ca\cb} = \Tg(\JT_\ca, \JT_\cb), \quad  \partial_{\xi_k} \ldots \partial_{\xi_1}\gtilde_{\ca\cb} = \TD_{\xi_k} \ldots \TD_{\xi_1}\Tg(\JT_\ca, \JT_\cb). \non
\ee
Expanding the right hand sides, all terms are bounded by previous results and hence the result follows.
\end{proof}

\section{The conformal extension theorem}

We have shown above how the boundedness of the tractor curvature derivatives to order $k$ leads to bounded components of the physical metric to $k^{th}$ order in $U$. If $V'=\phi(U)$, as in section \ref{nullcoord}, is a closed set then we can apply theorem 1 from \cite{Whitney1} to extend the metric components to a neighbourhood $\tilde{V}$ of $V$. However one of our conditions was that $\c$ be incomplete and hence $V$ is not closed as the ideal endpoint is not included. Therefore we need to use the main theorem from \cite{Whitney2}, which requires $V$ to satisfy the property $\mathcal{P}$. \footnote{A connected set $A$ satisfies $\mathcal{{P}} $ if there exists a positive constant $r$ such that for any two points $x, y$ with Euclidean distance $d$ in $\real^n$ there exists an arc in $A$ which connects $x, y$ and has length $d \times r$ or less. } If furthermore the $k^{th}$ derivatives of the metric components can be defined on the boundary of $V$ so that they are continuous on the closure of $V$ then one can find an extension for the metric.

The continuous extension of the function $f$ to the boundary is an important condition. It is not enough for $f$ to be bounded, as a simple example shows. The function $\sin (\frac{1}{r}) $ on $\real^n \setminus \{0\} $ is bounded but cannot be extended continuously to the origin along an isolated radial line, let alone in general. However if the function $f$ has bounded partial derivatives $f^{(k+1)}$ then its lower order derivatives $f^{(l)} ,(l \le k)$, are uniformly continuous and Lipschitz (see lemma 3.3.1 \cite{Racz}). Thus $f^{(k)} $ can be extended continuously to the boundary of $V$ and, provided property $\mathcal{P}$ holds, to $\real^n $. This is summarised in the following theorem:
\begin{theorem}
\label{conformalextthm}
Suppose we are given an incomplete conformal geodesics $\c$ with a congruence $C$ and a set $U$ as constructed above. Suppose $U $ satisfies the property $\mathcal{{P}} $. Let $\{e^i_\a\} $ and $\{ \TE^I_\TB\} $ denote the Weyl propagated vector and tractor frames.
If for $l \le k+1 $ the tractor curvature terms $\TD^l_e \TCurv\tensor{\mu\nu}{K}{L}$ have bounded norms in $U$ then the unphysical metric $ \gtilde_{ij} $ has a $C^k$-extension to $\tilde{U}=\psi(\tilde{V}) \supset U $.
\end{theorem}
If $\tilde{U} \setminus U$ contains points in $M$ then the extension is only a local extension as the extended metric may no longer be in the conformal class $[g]$.

\section{Conjugate points}\label{sectcjp}

So far we have assumed that our congruence is free of conjugate points. Now we derive some conditions that allow us to construct such a congruence.
We want to make sure that the separation vector $\eta^i $ between two neighbouring conformal geodesics doesn't vanish in $(M, \gtilde)$. However our congruence was constructed from metric null geodesics in $(M,g)$ where the Jacobi vectors are not the same as we are using two different velocities, respectively time coordinates along $C$. We use (\ref{tractorconfJac1}) to analyse $C$ for points where the deviation vector $\eta^i $ vanishes for $(M, \gtilde)$.
The Weyl propagated vectors $m^i, \mbar^i$ span a space-like 2-plane in each $T_{\c(\t)} M $. The projection onto this plane induces a positive definite metric $\hhat_{ij}= m_i \mbar_j + \mbar_j m_i $ with $\Dhat_v \hhat_{ij}=0$, where indices are raised an lowered with $\gtilde_{ij} $.

We observe the following equations
\bea
\label{Jacobitractorvtensor}
  \JT_Z :=\Tg(\JT,Z) = 0, \quad \JT_0:=\Tg(\JT, V)=\gtilde(\eta, v) \non \\
   \JT_\a:=\Tg(\JT,E_\a)=\gtilde(\eta, e_\a), \quad \JT_B:=\Tg(\JT, B)=\langle \btilde , \eta \rangle,
\eea
where $\btilde_i= b_i-\Ups_i$ is the 1-form of the congruence in $\Dtilde$. Then (\ref{Jacobitractorvtensor}) give rise to the following differential equations
\bea
\label{D2JT0}
  D \JT_0 &=& 0 \non \\
  D^2 \JT_{\app} &=& \TCurv(v, \eta)(E_\app, V) 
\non \\
  D^2 \JT_B &=& \TCurv(v, \eta)(B, V) 
\non  \\
  D^2 \JT_1 &=& \TCurv(v, \eta)(N, V) + 2 D \JT_B 
\non
\eea
It follows that $\JT_0$ is constant along the curves. Thus $\eta^i $ remains orthogonal to the conformal geodesic if $\JT_0$ vanishes initially. To analyse the behaviour of the components of the deviation vector $\eta^i $ projected into the 2-plane of $e_2^i, e_3^i $, we define $\hhat\tensor{i}{j}{}\eta^i=\eta^\app e^i_\app:=z \mu^i $ with $z^2:=\hhat(\eta,\eta)$ and $\hhat(\mu,\mu) =1$. It follows that $\Dhat_v(z\mu^i) $ lies in the 2-plane and
\be
  \ddot{z} = \frac{\hhat(\Dhat_v^2 \eta, \eta)}{z}+\frac{\hhat(\Dhat_v \eta,  \Dhat_v \eta)\hhat(\eta, \eta) - \hhat(\Dhat_v\eta, \eta)^2}{z^3} 
\non
\ee
Using the Cauchy-Schwartz inequality on the final two terms we find
\be
\label{zinequ}
  \ddot{z} \ge \frac{\hhat(\Dhat_v^2 \eta, \eta)}{z}
\ee
Suppose from now on that $\JT_0 = \gtilde(\eta,v) $ vanishes initially and hence along $\c$. Then $\eta^i=N_1 v^i + z \mu^i $ and hence
\be
\label{yforeta}
  \TCurv(v,\eta)\tensor{}{I}{J} = z \TCurv(v,\mu)\tensor{}{I}{J}
\ee
So for a deviation vector that starts in the $(e_2, e_3) $-subspace we have
\bea
  \hhat(D^2 \eta, \eta)
  &=& \eta_2 D^2\eta_2 + \eta_3 D^2\eta_3 \non \\
  &=& \TCurv(v,\eta)(E_2,V)\eta_2 + \TCurv(v,\eta)(E_3,V)\eta_3 \non \\
  &=& z^2 \TCurv(v,e_\bpp)(E_\app,V)\mu^{\app}\mu^{\bpp} \non \\
\label{TCurvinequ}
  &\ge & - \norm \TCurv \norm z^2 ,
\eea
where we used (\ref{yforeta}) in the last equality and the definition of the curvature norm for the inequality.

Let $\Lambda$ be a space-like 2-surface orthogonal to our congruence $C$ in $(M,g)$ and with $m^i, \mbar^i $ $\Dhat$-propagated and tangent to $\Lambda$. Similar to $\hhat_{ij}$ we can define the tractor metric $H_{IJ}=M_I \Mbar_J + \Mbar_I M_J $, which is positive definite for any tractor in the span of $M, \Mbar $, respectively $E_2, E_3 $, and satisfies $\TD_v H_{IJ}=0 $. We define
$$\chi_{iJ} = h\tensor{i}{k}{} \TD_k V^L H_{LJ} ,$$
which is a tractor analogue of the second fundamental form of $\Lambda$ (see appendix). We define the norm of $\chi_{iJ} $ as
$$
  \norm \chi_{iJ} \norm^2 = sup \{ H^{KL}\chi_{iK}\chi_{jL}\mu^i\mu^j : \hhat(\mu,\mu)=1\}.\non
$$
We can now prove the following theorem:
\begin{prop}
\label{noconjugatepointtheorem}
Suppose along the final segment of an incomplete null conformal geodesic $\c$ the tractor curvature has bounded norm, say $\norm \TCurv \norm < k^2$.
\bitem
\item Then two points p, q cannot be conjugate to one another along $\c$ unless their parameter distance is greater than or equal $\frac{\pi}{k}$;
\item Let $\Lambda $ be a space-like 2-surface, orthogonal to $\c$, used to construct the congruence $C$ and the null hypersurface $\Sigma$ above. Suppose the norm $\norm \chi_{iJ} \norm$  is bounded by $X$ at $p\in \c$. Then a point $q\in \c$ cannot be conjugate to $\Sigma$ along $\c$ unless the conformal parameter distance from $\Sigma$ is greater than or equal $T(k,X)=\frac{1}{k}\tan^{-1}(\frac{ k}{X}).$
\eitem
\end{prop}
This is the tractor analogue of corollaries 3.2.3 and 3.2.4 in \cite{Racz}.
\begin{proof}
Combining  (\ref{zinequ}, \ref{TCurvinequ}) and the boundedness condition we get $$\ddot{z} \ge - \norm \TCurv \norm z \ge -k^2 z.$$
First we show that a solution of (\ref{zinequ}) with initial data $z(0)=z_0 > 0, \dot{z}(0)=z_1 $ is positive as long as the solution of $\ddot{y} = -k^2 y$ with initial data $y(0)=z_0$, $\dot{y}(0)=z_1 $ has not vanished. In fact we obtain $z \ge y $ for this interval. The proof is the same as in \cite{article1}, setting $l=0$ there, and holds for the case $z_0=0, z_1>0$ as well.
Thus we have $$   z(\t) \ge z_0 \cos (k\t) +\frac{z_1}{k} \sin(k\t) $$ as long as the right hand side is positive.

For $z_1>0$ we can easily see that $z>0 $ for $\t \in (0,  \frac{\pi}{2k}) $ so that the result follows. In fact if we set $z_0=0, z_1=1$ then $z>0$ on $(0, \frac{\pi}{k})$. Thus any point $q$ conjugate to $p=\c(0)$ must lie outside this interval.

Now consider $z_1<0$. 
From $\JT_0=0$ and $\eta^i=N_1 v^i + z \mu^i  $ we deduce $z^2 = \hhat(\eta,\eta) = H(\JT,\JT)$. Then
$$Dz = \frac{H_{IJ}\JT^I \TD_v \JT^J }{z} =H_{IJ}\JT^I \TD_\mu V^J , $$
where we have used the tractor identities $\TD_v\JT^I= \TD_\eta V^I $ and $\TD_v V^I=0 $. Then the Cauchy-Schwarz inequality implies
$$
\mod Dz \mod \le z H_{IJ}\TD_\mu V^I \TD_\mu V^J \le z \norm \chi_{iJ} \norm. \non
$$
For $z_1<0$ we get $\frac{1}{X} \le -\frac{z_0}{z_1}$ and hence the result follows.

\end{proof}
The norm $\norm \chi_{iJ} \norm$ is effectively the norm defined in \cite{Racz} of the unphysical second fundamental form of $\Lambda$. It can be calculated from the physical second fundamental form and the initial data on $\Sigma$. For $\Lambda=exp_{(g)}[Q] $, $\chi_{ij}$ vanishes at $p$. If $b_k u^k =0$ then $X=0$ at $p$ and $T(k,X)=\frac{\pi}{2k}$.


Racz \cite{Racz} has shown how to construct an open neighbourhood $\mathcal{O} $ of $\c[(0, \t_F)] $ which satisfies the property $\mathcal{P} $. Thus we can state the following theorem for a local extension.
\begin{theorem}{\textbf{Local conformal extension theorem}}
\label{localextthm}

Suppose we are given an incomplete conformal geodesic $\c$ along which the tractor curvature has bounded norm. Then there exists a congruence $C$ and a set $U=\bar{\mathcal{O}}$ as constructed above, such that $U $ satisfies the property $\mathcal{{P}} $. 

If for $l \le k+1 $ the tractor curvature terms $\TD^l_e \TCurv\tensor{\mu\nu}{K}{L}$ have bounded norms in $U$ then the unphysical metric $ \gtilde_{ij} $ has a local $C^k$-extension to $\tilde{U}=\psi(\tilde{V}) \supset U $.
\end{theorem}
In \cite{CPEBH}, the conditions of theorem \ref{localextthm} are used to analyse the conformal properties of evaporating black hole models. These models are constructed by glueing Vaidya space-times between Schwarzschild at one end and flat space at the other end. The final point of the singularity is of particular interest. If it is approached from the past along an outgoing null geodesic no extension is possible, while strong assumptions on the matching conditions allow for a local conformal extension along a future-directed ingoing null geodesic.

\section{Conclusion and prospects}
In this article we have extended the work from \cite{article1} to null conformal geodesics. We have shown that, if sufficiently many derivatives of the tractor curvature have bounded norm, then we can find a conformal extension of $g_{ij}$ on a neighbourhood $U$ if it satisfies the property $\mathcal{P} $. The unphysical metric is derived directly from the null congruence and the initial data.

Note however that the extension in theorems \ref{conformalextthm} and \ref{localextthm} cannot be guaranteed to be global as neighbouring conformal geodesics might not be incomplete or remain in $U$. Suppose $U$ can be constructed such that all curves of $C$ are incomplete then one wants to know whether $U$ satisfies the property $\mathcal{P} $. Due to the construction of $U$ only the set $\partial M \cap \partial U$ can cause the property $\mathcal{P} $ to fail on $U$. Thus the structure of the singularity has to be taken into consideration. It remains to be investigated whether a space-like or null causal boundary might be sufficient to guarantee property $\mathcal{P} $ and hence a global extension.

The extension theorems may also be applied at null infinity. Take for example a vacuum space-time. We have seen above that the conformal parameter $\t$ is given by a M\"obius transformation acting on the affine parameter $s$. The initial data for $b_i$ can be chosen such that $\t$ is finite at null infinity and the space-time is conformally compactified. In asymptotically flat space-times one expects that the tractor curvature conditions are satisfied and the metric can be extended through null infinity. The problems of finding a global extension through conformal gauge singularities and null infinity will be considered elsewhere.

The derivation of an extension theorem for isotropic singularities was the main objective of this paper. However it was also used to study new tools for the initial value problem on isotropic singularities. It is hoped that similar to the study of the conformal structure in the vacuum case this will provide new approaches for numerical treatments of the problem. Thus we might get a bit closer to understanding how similar our universe is to the FLRW space-times and what the nature of the big bang is.

Finally I would like to thank Paul Tod and Mark Heinzle for their constructive criticism and suggestions. I would also like to thank ESI, where the research for this article was carried out.
\appendix
\section{The second fundamental form as a tractor}

We show how the tractor $\chi_{iJ} = h\tensor{i}{k}{} \TD_k V^L H_{LJ}$ defined for the space-like 2-surface $\Lambda$ is related to the second fundamental form of $\Lambda$ in the physical and unphysical space-time. We have $\Dhat = \D + b = \Dtilde + \btilde $ with $\btilde=b-\Ups$. In the $\Dtilde$-gauge the tensor $H\tensor{I}{J}{}$ takes the form
$$ H\tensor{I}{J}{} =   \left( \begin{array}{ccc} 0 & 0 & 0\\ \phi^j & \hhat\tensor{i}{j}{} & 0 \\ \theta  & \psi_i & 0\end{array} \right) \, where \quad \begin{array}{l}
\phi^j = \frac{1}{v^2} \btilde_i \hhat^{ij} \sect(\e_i[-2]), \\
\th\,\, = \frac{1}{v^2} \hhat^{ij} \btilde_i \btilde_j \sect(\e[-2]), \\
\psi_i = \btilde_i \hhat\tensor{}{i}{j} \sect(\e^j[0]), \end{array}$$
and indices on $\hhat_{ij} $ are lowered and raised with the unphysical metric. For the $\D$-gauge one just replaces $\btilde_i$ by $b_i$.

In the $\Dhat$-gauge and the $\Dtilde $-gauge $\TD_i V^J $ takes the following form:
$$
  \TD_i V^J
 = \vect{-v^{-1}\cg_{ik}v^k}{v^{-1}(\Dhat_i v^j + \btilde_i v^j)}{-v^{-1}\Phat_{ik}v^k}_{\Dhat}
 = \vect{-v^{-1}\cg_{ik}v^k}{v^{-1}\Dtilde_i v^j }{-v^{-1}\Ptilde_{ik}v^k}_{\Dtilde } \non
$$
The top and bottom entries are eliminated by the metrics and hence will not contribute to $\chi\tensor{i}{J}{}$. The unphysical second fundamental form of $\Lambda $ is given by  $\tilde{\chi}_{ij} = \hhat\tensor{i}{k}{}  \hhat\tensor{j}{l}{} \Dtilde_i v_j  = \hhat\tensor{i}{k}{}  \hhat\tensor{j}{l}{} \Dhat_i v_j  $. Thus
$$
  \chi\tensor{i}{J}{} = \vect{0}{v^{-1}\tilde{\chi}\tensor{i}{j}{}}{0}_{\Dhat }
= \vect{0}{v^{-1}\tilde{\chi}\tensor{i}{j}{}} {v^{-1}\tilde{\chi}\tensor{i}{j}{}\btilde_j}_{\Dtilde} \non
$$
so that $\norm \chi_{iJ} \norm$ is the norm of the unphysical second fundamental form as defined in \cite{Racz}.

Finally we give an expression of this norm in terms of the physical $\chi_{ij} $. The dyads for $\D $ and $ \Dhat$ are related by (\ref{spinormatrix}). Thus in the above formulae we replace $\hhat\tensor{i}{j}{}$ by $h\tensor{i}{j}{} = z_i \zbar^j + \zbar_i z^j$, whenever indices are in this position. 
Then $$ \norm \chi_{iJ} \norm = sup \{q^4 h^{pq}(\chi_{ip}+ \langle b,u \rangle h_{ip})(\chi_{jq}+ \langle b,u \rangle h_{jq})\nu^i \nu^j \,:\, h(\nu,\nu)=1  \}.$$
The additional $\langle b,u \rangle h_{ij}$ terms is the well-known correction term for the trace part under conformal rescaling, while the $q^4 $ term arises from rescaling $\hhat^{ij}$ and the normalisation condition on $\nu^i $.

Above the initial data was shown to be freely specifiable as long as it didn't introduce any poles into $\t $ or $b_i $ on the chosen final segment of $\c$. Thus we can use $\Lambda = exp_{(g)}Q $ with $q=1 $ and $\langle b,u \rangle = 0 $ on $\Sigma$ to make  $\norm \chi_{iJ}\norm $ vanish at $p$. If necessary we have to select a shorter final segment on $\c$.

\newpage
\providecommand{\bysame}{\leavevmode\hbox to3em{\hrulefill}\thinspace}
\providecommand{\href}[2]{#2}

\end{document}